\documentclass[12pt]{article}
\usepackage{amsmath}
\usepackage{amssymb, amscd, amsthm,amsfonts}
\usepackage{color}
\usepackage{xspace,colortbl}
\usepackage[all]{xy}
\usepackage[dvips]{graphicx}
\usepackage{verbatim}
\newtheorem{theo}[]{{\emph{Theorem}}}
\newtheorem{coro}[]{{\emph{Corollary}}}
\newtheorem{lemma}[]{{\emph{Lemma}}}

\theoremstyle{remark}
\newtheorem*{remark}{\textbf{Remark}}

\theoremstyle{definition}

\newcommand{\ra}{\rightarrow}

\newcommand{\calC}{\mathcal{C}}

\newcommand{\calF}{\mathcal{F}} 

\newcommand{\bF}{\mathbb{F}}

\newcommand{\cC}{\mathcal{C}}

\newcommand{\al}{\alpha}
\newcommand{\be}{\beta}
\newcommand{\ga}{\gamma}

\DeclareMathOperator{\Tra}{\mathrm Tr}

\begin{document}

\title{On Binary Cyclic Codes with Five Nonzero Weights} \maketitle

\begin{center}
$\mathrm{Jinquan\;\; Luo}$\footnote{J.Luo is with the School of
Mathematics, Yangzhou University, Jiangsu Province, 225009, China.
and also with the Division of Mathematical Sciences, School of
Physics and Mathematical Sciences, Nanyang Technological University,
Singapore. This work is supported by NRF Competitive Research
Program NRF-CRP2-2007-03, Singapore.}
\end{center}
\newpage
 \textbf{Abstract} \par Let $q=2^n$, $0\leq k\leq n-1$, $n/\gcd(n,k)$ be odd and $k\neq n/3, 2n/3$. In this paper
the value distribution of following exponential sums
\[\sum\limits_{x\in
\bF_q}(-1)^{\mathrm{Tr}_1^n(\alpha x^{2^{2k}+1}+\beta x^{2^k+1}+\ga
x)}\quad(\alpha,\beta,\ga\in \bF_{q})\] is determined. As  an
application, the weight distribution of the binary cyclic code
$\cC$, with parity-check polynomial $h_1(x)h_2(x)h_3(x)$ where
$h_1(x)$, $h_2(x)$ and $h_3(x)$ are the minimal polynomials of
$\pi^{-1}$, $\pi^{-(2^k+1)}$ and $\pi^{-(2^{2k}+1)}$ respectively
for a primitive element $\pi$ of $\bF_q$, is also determined.

\emph{Index terms:}\;Exponential sum, Cyclic code, Moment identity,
Weight distribution, Sequence
\newpage
\section{Introduction}

\quad Basic knowledge on finite fields could be found in \cite{Lid
Nie}. The following notations are fixed throughout this paper except
for specific statements.
\begin{itemize}
  \item Let $n$ be a positive integer, $q=2^n$,  $\bF_q$ be the finite field
  of order $q$. Let $\pi$ be a primitive element of $\bF_q$.
  \item Let $\Tra_i^j:\bF_{2^i}\ra\bF_{2^j}$ be the trace mapping,  and
$\chi(x)=(-1)^{\Tra_1^n(x)}$ be the canonical additive character on
$\bF_q$.
  \item Let $k$ be a positive integer, $1\leq k\leq n-1$ and  $k\notin \{\frac{n}{3},\frac{2n}{3}\}$.
  Let $d=\gcd(n,k)$, $q_0=2^d$, and $s=n/d$. Assume $s$ is odd.
\end{itemize}

 For cyclic code $\cC$ with length $l$, let $A_i$ be the
number of codewords in $\cC$ with Hamming weight $i$. The weight
distribution $\{A_0,A_1,\cdots,A_l\}$ is an important research
object for both theoretical and application interests in coding
theory. Classical coding theory reveals that the weight of each
codeword can be expressed by exponential sums so that the weight
distribution of $\cC$ can be determined if the corresponding
exponential sums can be calculated explicitly (see Feng and
Luo\cite{Fen Luo1}, Kasami\cite{Kasa1}, Luo and Feng\cite{Luo
Fen1}-\cite{Luo Fen2}, Moisio\cite{Mois},  van der
Vlugt\cite{Vand2}, Wolfmann\cite{Wolf}, Zeng, Liu and Hu\cite{Zen
Liu Hu}).

More precise speaking, let $q=2^n$, $\cC$ be the binary cyclic code
with length $l=q-1$ and parity-check polynomial
\[h(x)=h_1(x)\cdots h_u(x)\quad (u\geq 1)\]
where $h_i(x)$ $(1\leq i\leq u)$ are distinct irreducible
polynomials in $\bF_{2}[x]$ with the same degree $e_i$ $(1\leq i\leq
u)$, then $\mathrm{dim}_{\bF_{2}}\cC=\sum\limits_{i=1}^{u}e_i$. Let
$\pi^{-s_i}$ be a zero of $h_i(x)$, $1\leq s_i\leq q-2$ $(1\leq
i\leq u).$ Then the codewords in $\cC$ can be expressed by
\[c(\alpha_1,\cdots,\alpha_u)=(c_0,c_1,\cdots,c_{l-1})\quad (\alpha_1,\cdots,\alpha_u\in \bF_q)\]
where
$c_i=\sum\limits_{\lambda=1}^{u}\Tra^n_{1}(\alpha_{\lambda}\pi^{is_{\lambda}})$
$(0\leq i\leq l-1)$. Therefore the Hamming weight of the codeword
$c=c(\alpha_1,\cdots,\alpha_u)$ is {\setlength\arraycolsep{2pt}
\begin{equation} \label{Wei}
w_H\left(c\right)=2^{n-1}-\frac{1}{2}\,S(\alpha_1,\cdots,\alpha_u)
\end{equation}
} where
$f(x)=\alpha_1x^{s_1}+\alpha_2x^{s_2}+\cdots+\alpha_ux^{s_u}\in
\bF_{q}[x]$, $\bF_q^*=\bF_q\backslash\{0\}$,  and
\[S(\alpha_1,\cdots,\alpha_u)=\sum\limits_{x\in \bF_q}(-1)^{\Tra_1^n(\alpha_1x^{s_1}+\cdots+\alpha_ux^{s_u})}.\]
In this way, the weight distribution of cyclic code $\cC$ can be
derived from the explicit evaluating of the exponential sums
$S(\alpha_1,\cdots,\alpha_u)\;(\alpha_1,\cdots,\alpha_u\in \bF_q).$

 Let $h_1(x)$, $h_2(x)$ and $h_{3}(x)$
be the minimal polynomials of $\pi^{-1},\pi^{-(2^k+1)}$ and
$\pi^{-{(2^{2k}+1)}}$ over $\bF_{2}$ respectively. Then
$\mathrm{deg}\,h_i(x)=n\; \text{for}\; i=1,2,3.$
 Let
$\cC$ be the binary cyclic codes  with length $l=q-1$ and
parity-check polynomials $h_1(x)h_2(x)h_3(x)$. It is a consequence
that $\cC$ is the dual of the binary cyclic code whose defining
zeroes are $\pi^{2^{2k}+1}$ ,$\pi^{2^k+1}$ and $\pi$. Then the
dimensions of $\cC$ is $3n$ by excluding several special cases:
$k=n/3,2n/3$.

For $\al,\be,\ga\in \bF_q$,  define the exponential sum
\begin{equation}\label{def S}
S(\al,\be,\ga)=\sum\limits_{x\in \bF_q}\chi\left(\alpha
x^{2^{2k}+1}+\beta x^{2^k+1}+\ga x\right).
\end{equation}
 Then the complete
weight distributions of $\cC$ can be derived from the explicit
evaluation of $S(\al,\be,\ga)$.

For $k=1$, $\calC^{\bot}$ or its {\color{black}extended} code, is
the well-known triple-error correcting BCH code which has been
extensively studied.  For instance,
\begin{itemize}
  \item[(1).]
  The weight distribution of
  $\calC$ has been calculated, see MacWilliams and Sloane\cite{Mac
  Slo}, pp.669, Kasami\cite{Kasa1} for $n$ odd  and Berlekamp\cite{Berl} for $n$ even.
  \item[(2).] The covering radius of $\calC$ is 5, which has been
  proven in Assmus and Mattson\cite{Ass Mat}.
  \item[(3).]The coset
  distribution of $\calC^{\bot}$ has been determined, see Charpin,
  Helleseth and Zinoviev\cite{Cha Hel Zin2}.
  \item[(4).] The weight of coset leaders to the extended code of $\calC^{\bot}$ has been
studied,
  see Charpin,
  Helleseth and Zinoviev\cite{Cha Hel Zin1}-\cite{Cha Hel Zin3} and Charpin, Zinoviev\cite{Cha Zin}.
\end{itemize}

Let $a=\left(a_{\lambda}\right)_{\lambda=0}^{2^n-2}$ and
$b=\left(b_{\lambda}\right)_{\lambda=0}^{2^n-2}$ be two
$m$-sequences with period $q-2$. The \emph{correlation function} of
$a$ and $b$ for a shift $\tau$ is defined by

\[M_{{a},{b}}(\tau)=\sum\limits_{\lambda=0}^{2^n-2}(-1)^{a({\lambda})-b({\lambda+\tau})}\hspace{2cm}(0\leq \tau\leq
q-2).\]

Binary sequences with low cross correlation and auto-correlation are
widely used in Code Division Multiple Access(CDMA) spread spectrum
(see Simon, Omura and Scholtz\cite{Sim Omu}). Pairs of binary
$m$-sequences with few-valued correlations have been extensively
studied for several decades, see Dobbertin, Felke, Helleseth and
Rosendahl\cite{Dob Fel},   Helleseth\cite{Hell2}, Helleseth,
Kholosha and Ness\cite{Hel Kho}, Helleseth and Kumar \cite{Hel Kum},
Hu, Zeng, Li and Jiang\cite{Hu },  Johansen, Helleseth and
Tang\cite{Joh Hel Tan},
 Ness and Helleseth\cite{Nes
Hel2}-\cite{Nes Hel}, Niho\cite{Niho}, Rosendahl\cite{Rose}, Yu and
Gong\cite{Yu Gon1}-\cite{Yu Gon2} and references therein. Recently,
the exponential sum $S(\al,\be,0)$ with $n/d$ odd has been studied,
in terms of certain combination of exponential sums, see Johansen
and Helleseth\cite{Joh Hel}. In particular, for the case $k=1$ and
$n$ odd, the five-valued correlation distribution between two
$m$-sequences has been determined.

Based on the exponential sum $S(\al,\be,\ga)$, we could define  a
family of $m$-sequences $\calF=\calF_1\cup \calF_2$ with
\[
  \calF_1=\left\{\left(\Tra_1^n(\al \pi^{\lambda(2^{2k}+1)}+\be
  \pi^{\lambda(2^k+1)}+\pi^{\lambda})\right)_{\lambda=0}^{q-2}\Big{|}\,\al, \be\in \bF_{q} \right\}
\]
and
\[
  \calF_2=\left\{\left(\Tra_1^n(\al\pi^{\lambda(2^{2k}+1)}+
  \pi^{\lambda(2^k+1)})\right)_{\lambda=0}^{q-2}\Big{|}\,\be\in \bF_q \right\}\bigcup
  \left\{\left(\Tra_1^n(\pi^{\lambda(2^{2k}+1)})\right)_{\lambda=0}^{q-2}\right\}.
\]
The sequence family $\calF$ has family size $2^{2n}+2^n+1$.

 In this paper, we will focus on the case $n/d$ is odd and it is
presented as follows. In section 2 we introduce some preliminaries.
In section 3 we will determine the value distribution of
$S(\al,\be,\ga)$,  and at the same time, the weight distribution of
$\cC$. As a corollary, the possible correlation values among the
sequences in $\calF$ can be obtained. But unfortunately, we could
not determine the correlation distribution. The main techniques we
will employ are binary quadratic form theory and the third-power
moment identities of $S(\alpha,\beta,\ga)$.

\section{Preliminaries}

\quad We follow the notations in section 1. The first technique is
quadratic form theory over $\bF_{q_0}$.

 Let $H$
be an $s\times s$ matrix over $\bF_{q_0}$. For the quadratic form
\begin{equation}\label{qua for}
F:\bF_{q_0}^s\ra \bF_{q_0},\quad F(x)=XHX^T\quad
(X=(x_1,\cdots,x_s)\in \bF_{q_0}^s),
\end{equation}
define $r_F$ of $F$ to be the rank of the skew-symmetric matrix
$H+H^T$. Then $r_F$ is even.
 We have the following result on the exponential sum of binary quadratic forms (see \cite{Luo Tan Wan}).
\begin{lemma}\label{qua}

For the quadratic form $F=XHX^T$ defined in (\ref{qua for}),
\[\sum\limits_{X\in\bF_{q_0}^s}(-1)^{\Tra_1^{d}(F(X))}=\pm
q_0^{s-\frac{r_F}{2}}\; \text{or}\; 0\] Moreover, if $r_F=s$, then
\[\sum\limits_{X\in\bF_{q_0}^s}(-1)^{\Tra_1^{d}(F(X))}=\pm
q_0^{\frac{s}{2}}\]
\end{lemma}

The following result will be used in the study of $S(\al,\be,\ga)$
(see \cite{Vand2}).
\begin{lemma}\label{det gamma}
For the fixed quadratic form defined in  (\ref{qua for}), the value
distribution of
$\sum\limits_{X\in\bF_{q_0}^s}(-1)^{\Tra_1^d\left(F(X)+AX^T\right)}$
when $A$ runs through $\bF_{q_0}^s$ is shown as following
\[
\begin{array}{ccc}
value & \qquad\qquad\qquad&multiplicity \\[2mm]
0&\qquad\qquad\qquad&q_0^s-q_0^{r_F}\\[2mm]
q_0^{s-\frac{r_F}{2}}&\qquad\qquad\qquad &\frac{1}{2}(q_0^{r_F}+q_0^{\frac{r_F}{2}})\\[2mm]
-q_0^{s-\frac{r_F}{2}}&\qquad\qquad\qquad &\frac{1}{2}(q_0^{r_F}-q_0^{\frac{r_F}{2}})\\[2mm]
\end{array}
\]
\end{lemma}

Note that the field $\bF_q$ is a vector space over $\bF_{q_0}$ with
dimension $s$.For fixed basis $v_1,\cdots,v_s$ of $\bF_q$ over
$\bF_{q_0}$, each $x\in \bF_q$ can be uniquely expressed as
\[x=x_1v_1+\cdots+x_sv_s\quad (x_i\in \bF_{q_0}).\]
Thus we have the following $\bF_{q_0}$-linear isomorphism:
\[\bF_q\xrightarrow{\sim}\bF_{q_0}^s,\quad x=x_1v_1+\cdots+x_sv_s\mapsto
X=(x_1,\cdots,x_s).\] With this isomorphism, a function $f:\bF_q\ra
\bF_{q_0}$ induces a function $F:\bF_{q_0}^s\ra \bF_{q_0}$ where for
$X=(x_1,\cdots,x_s)\in \bF_{q_0}^s, F(X)=f(x)$ with
$x=x_1v_1+\cdots+x_sv_s$. In this way, function
$f(x)=\Tra_{d}^n(\gamma x)$ for $\gamma\in \bF_q$ induces a linear
form \begin{equation} F(X)=\Tra_{d}^n(\gamma
x)=\sum\limits_{i=1}^{s}\Tra_{d}^n(\gamma v_i)x_i=A_{\ga}X^T
\end{equation}\label{def A_gamma}
 where $A_{\ga}=\left(\Tra_{d}^n(\gamma
v_1),\cdots,\Tra_{d}^n(\gamma v_s)\right),$
 and
$f_{\alpha,\beta}(x)=\Tra_d^n(\alpha x^{p^{2k}+1}+\beta x^{p^k+1})$
for $\al,\be\in \bF_q$ induces a quadratic form
$F_{\alpha,\beta}(X)=XH_{\alpha,\beta}X^T$

From Lemma \ref{qua},  for $\alpha,\beta,\ga\in \bF_q$, in order to
determine the values of
\[S(\alpha,\beta,\ga)=\sum\limits_{x\in \bF_q}(-1)^{\Tra_1^n(\alpha
x^{2^{2k}+1}+\beta x^{2^k+1}+\ga x)}=\sum\limits_{X\in
\bF_{q_0}^s}(-1)^{\Tra_1^{d}\left(XH_{\alpha,\beta}X^T+A_{\ga}X^T\right)},\]
we need to determine the rank of $H_{\alpha,\beta}+H_{\al,\be}^T$
over $\bF_{q_0}$.

\begin{lemma}\label{rank}
For $\alpha,\beta\in \bF_q$ and $(\al,\be)\neq (0,0)$, let
$r_{\alpha,\beta}$ be the rank of
$H_{\alpha,\beta}+H_{\alpha,\beta}^T$.  Then the possible values of
$r_{\al,\be}$ are $s-1$ and $s-3$.
\end{lemma}
\begin{proof}
For $Y=(y_1,\cdots,y_s)\in \bF_{q_0}^s$, $y=y_1v_1+\cdots+y_sv_s\in
\bF_q$, we know that
\begin{equation}\label{bil form1}
F_{\alpha,\beta}(X+Y)-F_{\alpha,\beta}(X)-F_{\alpha,\beta}(Y)=2XH_{\alpha,\beta}Y^T
\end{equation}
is equal to
\begin{equation}\label{bil form2}
f_{\alpha,\beta}(x+y)-f_{\alpha,\beta}(x)-f_{\alpha,\beta}(y)=\Tra_{d}^n\left(y^{2^{2k}}(\alpha^{2^{2k}}
x^{2^{4k}}+\be^{2^{2k}} x^{2^{3k}}+\beta^{2^{k}} x^{2^{k}}+\al
x)\right)
\end{equation}

 Let
\begin{equation}\label{def phi}
\phi_{\al,\be}(x)=\alpha^{2^{2k}} x^{2^{4k}}+\be^{2^{2k}}
x^{2^{3k}}+\beta^{2^{k}} x^{2^{k}}+\al x.
\end{equation}
 Therefore,
\[{\setlength\arraycolsep{2pt}
\begin{array}{lcl}
r_{\al,\be}=r& \Leftrightarrow&\text{the number of common solutions of}\;XH_{\alpha,\beta}Y^T=0\;\text{for all}\;Y\in \bF_{q_0}^s\;\text{is}\; q_0^{s-r}, \\[2mm]
& \Leftrightarrow&\text{the number of common solutions of}\;\Tra_{d}^n\left(y^{2^{2k}}\cdot\phi_{\al,\be}(x)\right)=0\;\text{for all}\;y\in \bF_q\;\text{is}\; q_0^{s-r}, \\[2mm]
&\Leftrightarrow&\phi_{\al,\be}(x)=0\;\text{has}\; q_0^{s-r}\;
\text{solutions in}\; \bF_q.
\end{array}
}
\]

For a fixed algebraic closure $\bF_{2^\infty}$ of $\bF_2$, since the
degree of $2^{2k}$-linearized polynomial $\phi_{\al,\be}(x)$ is
$2^{4k}$ and $\phi_{\al,\be}(x)=0$ has no multiple roots in
$\bF_{2^{\infty}}$ (this fact follows from
$\phi'_{\al,\be}(x)=\al\in \bF_q^*$),  then the zeroes of
$\phi_{\al,\be}(x)$ in $\bF_{2^\infty}$, say $V$, form an
$\bF_{2^{k}}$-vector space of dimension 4.  Then $V\cap \bF_{2^n}$
is a vector space on $\bF_{2^{\gcd(n,k)}}=\bF_{2^{d}}$ of dimension
less that or equal to 4 since any elements in $\bF_{2^n}$ which are
linear independent over $\bF_{2^{d}}$ are also linear independent
over $\bF_{2^{k}}$(see \cite{Trac}, Lemma 4). Note that $s$ is odd
 and $r_{\al,\be}$ is always even. Hence the possible
  values of $r_{\al,\be}$ are $s-1$ and $s-3$.
\end{proof}

Another technique to determine the value distribution of
$S(\al,\be,\ga)$
 is the third-power moment identity of $S(\al,\be,\ga)$.
\begin{lemma}\label{moment}
For the exponential sum and $S(\al,\be,\ga)$, we have
\[
\sum\limits_{\al,\be,\ga\in
\bF_q}S(\al,\be,\ga)^3=(2^{n+d}+2^n-2^{d})\cdot 2^{3n}.
\]
\end{lemma}
\begin{proof}
We can calculate
\[
{ \setlength\arraycolsep{2pt}
\begin{array}{ll}
&\sum\limits_{\al,\be\in\bF_q}S(\al,\be,\ga)^3=\sum\limits_{x,y,z\in
\bF_q}\sum\limits_{\al\in
\bF_{q}}\chi\left(\al\left(x^{2^{2k}+1}+y^{2^{2k}+1}+z^{2^{2k}+1}\right)\right)\\[2mm]
&\qquad\sum\limits_{\be\in
\bF_q}\chi\left(\be\left(x^{2^k+1}+y^{2^k+1}+z^{2^k+1}\right)\right)\sum\limits_{\ga\in
\bF_q}\chi\left(\ga\left(x+y+z\right)\right)\\[2mm]
&\qquad=M_3\cdot 2^{3n}\end{array}}
\] where
$M_3$ is the number of solutions to the system of equations
\begin{eqnarray}\label{num L3}
\left\{
\begin{array}{ll}
  x+y+z=0&\\[2mm]
 x^{2^k+1}+y^{2^k+1}+z^{2^k+1}=0&\\[2mm]
x^{2^{2k}+1}+y^{2^{2k}+1}+z^{2^{2k}+1}=0
 \end{array}
 \right.
\end{eqnarray}

\begin{itemize}
  \item If $xyz=0$, we may assume $x=0$ and then $y=z$ which gives $2^n$ solutions. So is $y=0$ or $z=0$. Note that $x=y=z=0$ has been counted 3 times. Hence
   there are exactly $3\cdot 2^n-2$ solutions to (\ref{num L3}) satisfying $xyz=0$.
  \item If $xyz\neq 0$, then the number of solutions to (\ref{num L3}) is equal to $2^n-1$
  multiple of
  that to the system of equations
  \begin{equation}\label{red s3}
   x^{2^{2k}+1}+y^{2^{2k}+1}+1=x^{2^k+1}+y^{2^k+1}+1=x+y+1=0
  \end{equation}
  with $xy\neq 0$. By (\ref{red s3}) we have $x^{2^{2k}+1}+(x+1)^{2^{2k}+1}+1=x^{2^{k}+1}+(x+1)^{2^k+1}+1=0$
  which is equivalent to $x^{2^{k}}=x$.
  Hence $x\in \bF_{2^{k}}\cap \bF_{2^{n}}^*=\bF_{2^{d}}^*$ and $y=x+1$. Since $y\neq 0$, then $x\neq 1$. Therefore (\ref{red s3}) has
  $2^d-2$ solutions with $xy\neq 0$.

\end{itemize}
In total, we get $M_3=3\cdot 2^n-2+(2^d-2)(2^n-1)=2^{n+d}+2^n-2^d$.
\end{proof}

\section{The Weight Distribution of the Cyclic Code $\cC$}

\quad In the sequel we will give the the value distribution of
$S(\al,\be,\ga)$ and, at the same time, the weight distribution of
binary cyclic code $\cC$.

\begin{theo}\label{value dis S}
The value distribution of the multi-set
$\left\{S(\al,\be,\ga)\left|\al,\be,\ga\in \bF_q\right.\right\}$ and
the weight distribution of $\cC$ are shown as following (Column 1 is
the value of $S(\al,\be,\ga)$, Column 2 is the weight of
$c(\al,\be,\ga)$ and Column 3 is the corresponding multiplicity).

\begin{center}
\begin{tabular}{|c|c|c|}
\hline
value & weight & multiplicity \\[2mm]
\hline $2^{(n+d)/2}$ & $2^{n-1}-2^{(n+d-2)/2}$ &
$\frac{(2^{n-d-1}+2^{(n-d-2)/2})(2^n-1)(2^{n+2d}-2^{n}-2^{n-d}+2^{2d})}{2^{2d}-1}$
\\[2mm]
\hline $
-2^{(n+d)/2}$&$2^{n-1}+2^{(n+d-2)/2}$&$\frac{(2^{n-d-1}-2^{(n-d-2)/2})(2^n-1)(2^{n+2d}-2^{n}-2^{n-d}+2^{2d})}{2^{2d}-1}$
\\[2mm]
\hline
$2^{(n+3d)/2}$&$2^{n-1}-2^{(n+3d-2)/2}$&$\frac{(2^{n-3d-1}+2^{(n-3d-2)/2})(2^{n-d}-1)(2^n-1)}{2^{2d}-1}$
\\[2mm]
\hline
$-2^{(n+3d)/2}$&$2^{n-1}+2^{(n+3d-2)/2}$&$\frac{(2^{n-3d-1}-2^{(n-3d-2)/2})(2^{n-d}-1)(2^n-1)}{2^{2d}-1}$
\\[2mm]
\hline
$0$&$2^{n-1}$&$\scriptstyle{(2^n-1)(2^{2n}-2^{2n-d}+2^{2n-4d}+2^n-2^{n-d}-2^{n-3d}+1)}$
\\[2mm]
\hline $2^n$&$0$&$1$
\\[2mm]
\hline
\end{tabular}
\end{center}

\end{theo}
\begin{proof}
Let $n_i$ to be the number of pairs $(\al,\be)\in
\bF_q\backslash\{(0,0)$ such that $r_{\al,\be}=s-i$. Define
\[\Xi=\left\{(\al,\be,\ga)\in \bF_q^3\left|S(\al,\be,\ga)=0 \right.\right\}\]
and $\xi=\big{|}\Xi\big{|}$.

Since $n/d$ is odd, from Lemma \ref{det gamma}, Lemma \ref{rank} and
Lemma \ref{moment} we have
\begin{equation}\label{n/d odd 1}
n_1+n_3=2^{2n}-1
\end{equation}

\begin{equation}\label{n/d odd 2}
n_1+2^{2d}\cdot n_3=2^{n-d}(2^d+1)(2^n-1).
\end{equation}

These two equations yield
\begin{equation}\label{value n1 n3}n_1=\frac{(2^{n}-1)(2^{n+2d}-2^n-2^{n-d}+2^{2d})}{2^{2d}-1},\quad
n_3=\frac{(2^{n-d}-1)(2^n-1)}{2^{2d}-1}\end{equation}

Note that $S(0,0,\ga)$=0 unless $\ga=0$.  From Lemma \ref{det gamma}
we get that
\begin{equation}\label{value xi1}
{\setlength\arraycolsep{2pt}
\begin{array}{lll}
\xi&=2^{n}-1+(2^{n}-2^{n-d})n_{1}+(2^{n}-2^{n-3d})n_{3}&\\[2mm]
&=(2^n-1)(2^{2n}-2^{2n-d}+2^{2n-4d}+2^n-2^{n-d}-2^{n-3d}+1)
\end{array}
}
\end{equation}
Then the result follows from (\ref{value n1 n3}), (\ref{value xi1})
by using Lemma \ref{det gamma}.
\end{proof}

We hereby could give the possible values of correlation function
among sequences in $\calF$. For example, let
$a_{\al_1,\be_1}=\left(\Tra_1^n(\al_1 \pi^{\lambda(2^{2k}+1)}+\be_1
  \pi^{\lambda(2^k+1)}+\pi^{\lambda})\right)_{\lambda=0}^{q-2}$ and $a_{\al_2,\be_2}=\left(\Tra_1^n(\al_2 \pi^{
  \lambda(2^{2k}+1)}+\be_2
  \pi^{\lambda(2^k+1)}+\pi^{\lambda})\right)_{\lambda=0}^{q-2}$. Then the correlation function of $a_{\al_1,\be_1}$ and
  $a_{\al_2,\be_2}$ by a shift $\tau$
($0\leq \tau\leq q-2$) is
\[
\begin{array}{ll}
&C_{(\al_1,\be_1),(\al_2,\be_2)}(\tau)=\sum\limits_{\lambda=0}^{q-2}(-1)^{a_{\al_1,\be_1}({\lambda})-
a_{\al_2,\be_2}({\lambda+\tau})}\\[2mm]
&\qquad =\sum\limits_{\lambda=0}^{q-2}(-1)^{\Tra_1^n(\al_1
  \pi^{\lambda(2^{2k}+1)}+\be_1
  \pi^{\lambda(2^k+1)}+\pi^{\lambda})-\Tra_1^n(\al_2 \pi^{(\lambda+\tau)(2^{2k}+1)}+\be
  \pi^{(\lambda+\tau)(2^k+1)}+\pi^{\lambda+\tau})}\\[2mm]
  &\qquad = S(\al',\be',\ga')-1
  \end{array}
\]
 where
 \begin{equation}\label{coe cor}
 \al'=\al_1-\al_2 \pi^{\tau(2^{2k}+1)},\quad
 \be'=\be_1-\be_2\pi^{\tau(2^k+1)},\quad \ga'=1-\pi^{\tau}.
 \end{equation}
 \begin{remark}
 \end{remark}
As a corollary, we have

\begin{coro}
The non-trivial correlation values of the sequences in $\calF$ is
$-1$, $\pm 2^{\frac{n+d}{2}}-1$ and $\pm 2^{\frac{n+3d}{2}}-1$.
\end{coro}

\section{Conclusion and Further Study}

\quad In this paper we have studied the exponential sums
 $S(\al,\be,\ga)$ with $\al,\be,\ga\in \bF_{2^n}$.
After giving the value distribution of $S(\al,\be,\ga)$, we
determine the weight distributions of the cyclic codes $\cC$.

For the case $n/d$ even, we could get the possible values of
$S(\al,\be,0)$ and $S(\al,\be,\ga)$. But the first four moment
identities $\sum\limits_{\al,\be\in \bF_q}S(\al,\be,0)^i$ and
$\sum\limits_{\al,\be\in \bF_q}S(\al,\be,\ga)^i$ for $0\leq i\leq 3$
is not enough to determine the value distribution of $S(\al,\be,0)$
and $S(\al,\be,\ga)$. However, we could get the possible non-trivail
weights of $\calC$: $2^{n-1}$, $2^{n-1}\pm 2^{\frac{n}{2}+d-1}$ and
$2^{n-1}\pm 2^{\frac{n}{2}+2d-1}$. New machinery and technique
should be proposed to attack this problem.

\section{Acknowledgements}
\quad The authors will thank the anonymous referees for their
helpful comments.

\end{document}